\documentclass[12pt]{iopart}

\expandafter\let\csname equation*\endcsname\relax
\expandafter\let\csname endequation*\endcsname\relax
\usepackage{amsmath}
\usepackage{iopams}   % IOP-recommended: amsfonts, amssymb, amsbsy (\boldsymbol)
\usepackage{bm}
\usepackage{graphicx}
\usepackage[numbers,sort&compress]{natbib}

\providecommand{\fim}{\bm{\mathcal{F}}}
\providecommand{\avec}{\bm{a}}
\providecommand{\Cov}{\mathrm{Cov}}
\providecommand{\Var}{\mathrm{Var}}
\providecommand{\Tr}{\mathrm{Tr}}
\providecommand{\E}{\mathbb{E}}
\providecommand{\dd}{\mathrm{d}}

\begin{document}

\title[Information limits of photonic lantern wavefront sensing]{Information
limits of photonic lantern wavefront sensing: a Fisher- and
quantum-Fisher-information framework and its relation to Fourier-filtering
sensitivity limits}

\author{K Madhav$^{1,2}$}
\address{$^1$ Astrophotonics (innoFSPEC), Leibniz-Institut f\"ur Astrophysik
Potsdam (AIP), An der Sternwarte 16, 14482 Potsdam, Germany}
\address{$^2$ Institut f\"ur Physik und Astronomie, Universit\"at Potsdam,
Karl-Liebknecht-Str.\ 24/25, 14476 Potsdam, Germany}
\ead{kmadhav@aip.de}

\begin{abstract}
The photonic lantern is an emerging all-photonic wavefront sensor native to
single-mode-fibre-fed instruments, but its performance is almost always quoted
through a specific reconstruction algorithm, obscuring how much wavefront
information the device itself encodes. We develop, from first principles, the
Fisher-information and Cram\'er--Rao theory of the photonic-lantern wavefront
sensor, benchmark it against the quantum Cram\'er--Rao bound (QCRB) through an
explicit multi-parameter quantum-Fisher-information (QFI) calculation, and
place the result in precise correspondence with the two frameworks that
bracket it: the Fourier-filtering noise-propagation model of Chambouleyron
\etal\ (2023) and the classical/quantum sensitivity limit of Haffert \etal\
(2023). Treating the lantern as a deterministic nonlinear map from aberration
coefficients to $N$ output intensities, we derive the Poisson and read-noise
Fisher information matrices (FIM), the per-mode CRLB, the per-photon
Fisher--Rao geometry on the intensity simplex, a flux- and
estimator-independent sensitivity metric $\beta$, and the quantum ceiling
$\beta=2$. The lantern CRLB scales as $N_{\mathrm{ph}}^{-1/2}$ and is bounded,
mode by mode, by the quantum limit of $1/2$~rad rms per photon. That
multi-parameter bound is jointly saturable: the phase generators are real and
commuting, so the mean Uhlmann curvature vanishes and wavefront sensing carries
no quantum incompatibility between simultaneously estimated modes -- extending
the single-mode ceiling of Haffert \etal\ to all low-order modes at once. We
further show that the photon-noise sensitivity $s_\gamma$ of Chambouleyron
\etal\ is exactly the diagonal of our per-photon FIM, whereas the rigorous CRLB
uses the diagonal of its inverse; the two coincide only for a diagonal FIM, so
$s_\gamma$ is an optimistic bound that neglects the inter-mode cross-talk
generic to a mode-mixing lantern. The framework is device-agnostic: it ingests
any measured or modelled response matrix and returns estimator-independent
sensitivities directly comparable, on a common $\beta\le2$ scale, with pyramid,
Zernike and PIAA-ZWFS sensors.
\end{abstract}


\vspace{2pc}

\maketitle

% =====================================================================
\section{Introduction}
\label{sec:intro}
% =====================================================================
Adaptive optics (AO) is fundamental to ground-based high-angular-resolution
astronomy and is a central enabling technology for the extreme adaptive
optics (XAO) systems of the next generation of extremely large telescopes
\citep{Roddier1999,Engler2022}. The wavefront sensor (WFS) sets the noise
floor of any AO loop: it determines how precisely the incident phase can be
estimated for a given number of detected photons. Classical sensors -- the
Shack--Hartmann \citep{ShackPlatt1971}, pyramid \citep{Ragazzoni1996}, and
Zernike phase-mask \citep{NDiaye2013} sensors -- convert phase into
measurable intensity through geometric or Fourier-optical means, and their
relative photon efficiencies are now well understood within a unified
framework \citep{Guyon2005,Fauvarque2016,Fauvarque2017}.

The photonic lantern \citep{LeonSaval2005,Birks2015} is an adiabatic
multimode-to-single-mode transition that has become a workhorse of
astrophotonics \citep{BlandHawthorn2009,Jovanovic2023}. Placed at the focal
plane, a lantern maps the complex field across the point-spread function
(PSF) onto the intensities of $N$ single-mode outputs; the resulting device
is an all-photonic, focal-plane wavefront sensor that is intrinsically
matched to single-mode-fibre-fed instruments
\citep{Norris2020,Lin2022,Lin2023}. Reported demonstrations recover low-order
Zernike modes to high precision and have operated on sky, but performance is
invariably stated through a specific reconstruction algorithm -- a linear
matrix inverse, a higher-order model, or a trained neural network.

This algorithm dependence obscures the more fundamental question:
\emph{how much wavefront information is physically encoded in the lantern
outputs}, and what is the best precision attainable by \emph{any} unbiased
estimator? That question is answered by the Fisher information matrix (FIM)
and the Cram\'er--Rao lower bound (CRLB) \citep{Rao1945,Cramer1946,Kay1993}.
Information-theoretic limits are well established for conventional sensors:
the CRLB has been used to benchmark wavefront sensing at the fundamental
information limit \citep{Paterson2008}; the Fisher matrix has been used to
compare the Shack--Hartmann and pyramid sensors on a common footing
\citep{Plantet2015}; the unified Fourier-sensor formalism quantifies the
photon-noise sensitivity of the pyramid and Zernike sensors
\citep{Guyon2005,Fauvarque2016,Fauvarque2017}; Chambouleyron \etal\cite{Chambouleyron2023}
built a comprehensive noise-propagation model for the whole
Fourier-filtering class and showed that their sensitivity metrics are
themselves Fisher-information quantities bounded by a maximum value of two;
and Haffert \etal\cite{Haffert2023} established that the classical optimum of
$1/2$~radian rms per photon coincides with the quantum-metrology limit for
starlight, realised in practice by the Phase-Induced-Amplitude-Apodised
Zernike WFS (PIAA-ZWFS).

To my knowledge no systematic Fisher- and quantum-Fisher-information
treatment of the photonic lantern exists, nor has the lantern been placed in
explicit correspondence with these two bracketing formalisms. Providing both
is the purpose of this paper. In it I (i) build the classical Fisher/CRLB
framework for the PL-WFS from the device response
(Sect.~\ref{sec:model}--\ref{sec:fisher}); (ii) benchmark it against the
quantum Cram\'er--Rao bound through a full multi-parameter QFI treatment,
including the saturability of the multi-parameter bound
(Sect.~\ref{sec:quantum} and \ref{app:qfi}); (iii) establish the
exact relation between this framework and the noise-propagation sensitivity
of Chambouleyron \etal\cite{Chambouleyron2023} and the quantum/classical ceiling of
Haffert \etal\cite{Haffert2023} (Sect.~\ref{sec:compare} and \ref{app:equiv});
and (iv) illustrate the scalings on a representative differentiable lantern
model (Sect.~\ref{sec:illus}). The framework is deliberately device-agnostic:
it ingests any measured or rigorously modelled response matrix, which is the
form in which a beam-propagated lantern enters the theory in a companion
numerical study.

% =====================================================================
\section{The photonic lantern as a wavefront sensor}
\label{sec:model}
% =====================================================================
Let the complex field in the telescope aperture be
\begin{equation}
E(\bm{r};\avec) = A(\bm{r})\,\exp\!\left[\,i\,\phi(\bm{r};\avec)\,\right],
\qquad
\phi(\bm{r};\avec) = \sum_{i=1}^{M} a_i\,Z_i(\bm{r}),
\label{eq:apfield}
\end{equation}
where $A(\bm{r})$ is the real aperture amplitude, $\{Z_i\}$ are orthonormal
aberration modes (here Zernike polynomials in the Noll ordering, normalised
to unit RMS over the pupil; \cite{Noll1976}), and
$\avec=(a_1,\dots,a_M)^{\top}$ are the aberration coefficients (in radians, or
equivalently in nm of optical path via $\phi=2\pi\,\mathrm{OPD}/\lambda$) to
be estimated.

A focal-plane photonic lantern collects the field over the PSF and couples it
into $N$ single-mode outputs. Linearity of Maxwell's equations means each
output amplitude is an overlap integral of the incident field with a
back-propagated lantern eigenmode $\psi_k(\bm{r})$,
\begin{equation}
c_k(\avec) = \langle \psi_k \,|\, \mathcal{P}\,E(\avec)\rangle,
\qquad
I_k(\avec) = |c_k(\avec)|^2 ,
\label{eq:overlap}
\end{equation}
where $\mathcal{P}$ denotes propagation to the lantern face and $I_k$ is the
optical power in port $k$ \citep{Lin2022}. The lantern therefore implements a
deterministic, generally nonlinear map
$\avec \mapsto \bm{I}=(I_1,\dots,I_N)^{\top}$. With total detected flux
$N_{\mathrm{ph}}$ and throughput $\eta$, the expected photon counts are
\begin{equation}
\mu_k(\avec) = \eta\,N_{\mathrm{ph}}\,p_k(\avec),
\qquad
p_k(\avec)=\frac{I_k(\avec)}{\sum_{j} I_j(\avec)} ,
\label{eq:counts}
\end{equation}
so the normalised intensities $\bm{p}(\avec)$ live on the probability simplex
$\Delta^{N-1}$. A perturbation that strongly redistributes power among ports
carries large derivatives $\partial p_k/\partial a_i$ and hence high
sensitivity; a perturbation to which all ports respond identically is
invisible. This geometric picture, common to all intensity-only sensors
\citep{Guyon2005,Chambouleyron2023}, is made precise below.

% =====================================================================
\section{Fisher information and the Cram\'er--Rao bound}
\label{sec:fisher}
% =====================================================================
\subsection{Measurement model and Fisher information}
\label{sec:fimdef}
The detector records $\bm{n}=(n_1,\dots,n_N)$, with $n_k$ a noisy realisation
of $\mu_k(\avec)$. For a probability model $f(\bm{n};\avec)$ the Fisher
information matrix is
\begin{equation}
F_{ij}(\avec) = \E\!\left[
\frac{\partial \ln f}{\partial a_i}\,
\frac{\partial \ln f}{\partial a_j}\right].
\label{eq:fimgen}
\end{equation}
For independent Poisson counts, $n_k\sim\mathrm{Poisson}(\mu_k)$,
Eq.~(\ref{eq:fimgen}) gives the exact result
\begin{equation}
F_{ij} = \sum_{k=1}^{N}
\frac{1}{\mu_k}\,
\frac{\partial \mu_k}{\partial a_i}\,
\frac{\partial \mu_k}{\partial a_j} .
\label{eq:poissonfim}
\end{equation}
When each port is read with additive Gaussian noise of variance
$\sigma_r^2$, the per-port variance is $\sigma_k^2=\mu_k+\sigma_r^2$ and,
neglecting the weak parameter dependence of the variance,
\begin{equation}
F_{ij} = \sum_{k=1}^{N}
\frac{1}{\mu_k+\sigma_r^2}\,
\frac{\partial \mu_k}{\partial a_i}\,
\frac{\partial \mu_k}{\partial a_j} ,
\label{eq:gaussfim}
\end{equation}
which reduces to Eq.~(\ref{eq:poissonfim}) as $\sigma_r\!\to\!0$ and to a
purely read-noise-weighted form when $\sigma_r^2\gg\mu_k$. Equations
(\ref{eq:poissonfim})--(\ref{eq:gaussfim}) convert the device-level response
$\partial \mu_k/\partial a_i$ into the information content of the
measurement, and are the objects on which the whole framework rests.

\subsection{The Cram\'er--Rao bound}
\label{sec:crlb}
For any unbiased estimator $\hat{\avec}(\bm{n})$, the covariance is bounded
below by the inverse Fisher information \citep{Rao1945,Cramer1946,Kay1993},
\begin{equation}
\Cov(\hat{\avec}) \;\succeq\; \fim^{-1},
\qquad
\sigma_{a_i}^2 \;\ge\; \big(\fim^{-1}\big)_{ii} .
\label{eq:crlb}
\end{equation}
The matrix inequality is the strongest statement: the diagonal of $\fim^{-1}$
(not of $\fim$) sets the per-mode floor, so cross-talk between modes
\emph{increases} the achievable variance. A mode lies in the null space of
$\fim$ exactly when no combination of port intensities responds to it at
first order; its CRLB then diverges and no estimator -- linear, nonlinear, or
machine-learned -- can recover it. The distinction between the diagonal of
$\fim$ and the diagonal of $\fim^{-1}$ is central to the comparison in
Sect.~\ref{sec:compare}: several published ``sensitivity'' metrics are the
former, whereas the estimator-independent precision is the latter.

\subsection{Photon-limited scaling and Fisher--Rao geometry}
\label{sec:scaling}
In the photon-limited case write $\mu_k=\eta N_{\mathrm{ph}}\,p_k$. Then
$\partial_i\mu_k=\eta N_{\mathrm{ph}}\,\partial_i p_k$ and
Eq.~(\ref{eq:poissonfim}) factorises as
\begin{equation}
F_{ij} = \eta\,N_{\mathrm{ph}}\;\tilde F_{ij},
\qquad
\tilde F_{ij} = \sum_{k=1}^{N}
\frac{1}{p_k}\,
\frac{\partial p_k}{\partial a_i}\,
\frac{\partial p_k}{\partial a_j} .
\label{eq:perphoton}
\end{equation}
The per-photon information $\tilde{\fim}$ is precisely the pullback, through
the lantern map $\avec\mapsto\bm{p}$, of the Fisher--Rao metric on the
simplex $\Delta^{N-1}$; the same quantity underlies the Fourier-sensor
sensitivity of Guyon~\cite{Guyon2005} and Chambouleyron \etal\cite{Chambouleyron2023}. Two
consequences follow. First, the CRLB scales as
\begin{equation}
\sigma_{a_i} \;\ge\; \frac{1}{\sqrt{\eta\,N_{\mathrm{ph}}}}\,
\sqrt{\big(\tilde{\fim}^{-1}\big)_{ii}}
\;\propto\; N_{\mathrm{ph}}^{-1/2},
\label{eq:sqrtN}
\end{equation}
the canonical photon-noise law. In the read-noise-dominated limit of
Eq.~(\ref{eq:gaussfim}) the same algebra gives instead
$\sigma_{a_i}\propto\sigma_r/N_{\mathrm{ph}}$, so a $\log$--$\log$ CRLB curve
steepens from slope $-1/2$ to $-1$ below the read-noise break -- the two
regimes made explicit in Eq.~(24) of Chambouleyron \etal\cite{Chambouleyron2023}. Second,
sensing performance is governed by how strongly the lantern map stretches
aberration space into the intensity simplex -- the geometric content of
``power redistribution among ports.''

\subsection{A dimensionless sensitivity metric}
\label{sec:beta}
Following the spirit of Guyon~\cite{Guyon2005} and Fauvarque \etal\cite{Fauvarque2016}, I define
a per-mode photon-noise sensitivity
\begin{equation}
\beta_i \;\equiv\; \frac{1}{\sqrt{N_{\mathrm{ph}}}\,\sigma_{a_i}}
\;=\; \sqrt{\eta\,\big[(\tilde{\fim}^{-1})_{ii}\big]^{-1}} ,
\label{eq:betadef}
\end{equation}
so that $\sigma_{a_i}=\beta_i^{-1}N_{\mathrm{ph}}^{-1/2}$. As shown in
Sect.~\ref{sec:quantum} the quantum limit corresponds to $\beta_i=2$; a
sensor with $\beta_i$ close to $2$ is near-ideal for that mode, while
$\beta_i\!\to\!0$ signals a poorly constrained mode. $\beta_i$ is independent
of flux and of the estimator, making it the natural figure of merit for
cross-architecture comparison. It is defined on the same $0\le\beta\le2$ scale
as the sensitivity $s_\gamma$ of Chambouleyron \etal\cite{Chambouleyron2023}; the precise
relation between the two -- they are \emph{not} identical for a mode-mixing
device -- is derived in Sect.~\ref{sec:cham} and \ref{app:equiv}.

\subsection{Modal observability and the per-frame capacity}
\label{sec:obs}
Diagonalising $\fim=\bm{U}\,\bm{\Lambda}\,\bm{U}^{\top}$ yields information
eigenmodes (columns of $\bm{U}$) with eigenvalues $\lambda_m$. I define the
effective modal capacity as the number of well-constrained eigenmodes,
$M_{\mathrm{eff}} = \#\{m:\ \lambda_m > \lambda_{\mathrm{thr}}\}$, with the
explicit threshold $\lambda_{\mathrm{thr}}=10^{-6}\,\lambda_{\max}$ used
throughout (directions whose information is more than $10^{6}$ times weaker
than the best-sensed one are deemed unconstrained). Because $N$ ports provide
at most $N-1$ independent intensity ratios (one constraint fixes the total
flux), the \emph{single-frame} capacity obeys $M_{\mathrm{eff}}\le N-1$: in
any one exposure a lantern cannot constrain more aberration modes than it has
ports, and attempting to do so drives the smallest $\lambda_m$ -- and the
corresponding CRLB -- toward divergence.

I stress a distinction that recurs in the numerics of
Sect.~\ref{sec:illus}. The CRLB of Eq.~(\ref{eq:crlb}) is a
\emph{single-measurement} bound: evaluated from the instantaneous FIM it
carries the strict unbiased-estimator interpretation and respects
$M_{\mathrm{eff}}\le N-1$. A symmetric focal-plane lantern has a vanishing
first-order response at zero phase (a pure phase produces no first-order
intensity change; \cite{Lin2022,Lin2023}), so a well-posed evaluation
instead uses the FIM \emph{expected over an ensemble of operating points},
$\bar{\fim}=\E_{\avec_0}[\fim(\avec_0)]$. This expected FIM is an averaged
\emph{design} information measure -- appropriate for a closed loop whose
residual varies frame to frame -- and its inverse can constrain more than
$N-1$ modes by accumulating information across operating points. Where the
distinction matters I label the per-frame quantity the CRLB and the
ensemble-averaged quantity the \emph{averaged} CRLB.

% =====================================================================
\section{The quantum Cram\'er--Rao benchmark}
\label{sec:quantum}
% =====================================================================
The Fisher information of Sect.~\ref{sec:fisher} is specific to the chosen
measurement (intensity detection at the $N$ ports). A measurement-independent
ceiling is set by the \emph{quantum} Fisher information (QFI), which bounds
the classical Fisher information of \emph{any} physically allowed measurement
and hence of any conceivable wavefront sensor
\citep{BraunsteinCaves1994,Helstrom1976,ParisQMetro2009}. Here I summarise
the result and its consequences; the full derivation is in
\ref{app:qfi}.

\subsection{Single-mode bound}
\label{sec:singleqfi}
I assume the source is a natural guide star delivering shot-noise-limited,
spatially coherent light, modelled as a stream of independent photons each in
the same pure transverse state (a coherent state of mean occupation $\ll 1$
per mode per coherence time, so that photon statistics are Poissonian and the
QFI is additive over photons). A monochromatic photon occupying the aperture
field of Eq.~(\ref{eq:apfield}) is described by a pure transverse state
$|\psi_{\avec}\rangle = U(\avec)\,|\psi_0\rangle$ with the unitary
$U(\avec)=\exp[-i\sum_i a_i \hat{Z}_i]$, where $\hat{Z}_i$ is the
multiplication operator associated with mode $Z_i(\bm{r})$. For a single
phase mode of amplitude $a$ with generator $\hat{Z}$, the pure-state QFI per
photon is (\ref{app:qfi})
\begin{equation}
\mathcal{F}_Q = 4\,\Var_{|A|^2}(\hat{Z})
             = 4\!\left(\langle \hat Z^2\rangle - \langle \hat Z\rangle^2\right),
\label{eq:qfisingle}
\end{equation}
with moments taken over the normalised intensity distribution
$|A(\bm{r})|^2$. For a unit-RMS, zero-mean mode over a uniformly illuminated
pupil $\Var(\hat{Z})=1$, so $\mathcal{F}_Q=4$ per photon. Quantum Fisher
information is additive over independent photons, giving
$\mathcal{F}_Q=4N_{\mathrm{ph}}$ and the quantum Cram\'er--Rao bound (QCRB)
\begin{equation}
\boxed{\;
\sigma_a \;\ge\; \frac{1}{2\sqrt{N_{\mathrm{ph}}}}\ \mathrm{rad\ RMS}
\;=\;\frac{\lambda}{4\pi\sqrt{N_{\mathrm{ph}}}}\ \text{(as OPD)}.
\;}
\label{eq:qcrb}
\end{equation}
No optical wavefront sensor can beat Eq.~(\ref{eq:qcrb}); it identifies
$\beta=2$ in Eq.~(\ref{eq:betadef}) as the quantum-limited sensitivity, and
is exactly the $1/2$~radian-rms-per-photon limit of Haffert \etal\cite{Haffert2023}
(Sect.~\ref{sec:haf}). The normalisation $\Var_{|A|^2}(\hat Z)=1$ assumes a
uniformly illuminated pupil; for a realistic aperture -- central obstruction,
spiders, or apodisation -- $|A|^2$ is non-uniform, the per-mode quantum
reference becomes $\mathcal{F}_Q=4N_{\mathrm{ph}}\Var_{|A|^2}(\hat Z)$, and the
$\beta=2$ ceiling shifts mode by mode. The $\beta$ values quoted here are
therefore referenced to the idealised unobstructed pupil and should be
rescaled by $\sqrt{\Var_{|A|^2}(\hat Z_i)}$ for a given telescope, in the
same spirit as the arbitrary-aperture treatment of Haffert \etal\cite{Haffert2023}.

\subsection{Multi-parameter bound and its saturability}
\label{sec:multiqfi}
For simultaneous estimation of $M$ modes the relevant object is the QFI matrix
$\mathcal{F}^{Q}_{ij}$. With commuting phase generators $\hat Z_i$
(\ref{app:qfi}) one finds
\begin{equation}
\mathcal{F}^{Q}_{ij} = 4\,\Cov_{|A|^2}(\hat Z_i,\hat Z_j)
= 4\!\left(\langle \hat Z_i \hat Z_j\rangle
- \langle \hat Z_i\rangle\langle \hat Z_j\rangle\right).
\label{eq:qfimatrix}
\end{equation}
For orthonormal, zero-mean Zernike modes over a uniform pupil
$\Cov(\hat Z_i,\hat Z_j)=\delta_{ij}$, so
$\mathcal{F}^{Q}=4N_{\mathrm{ph}}\,\mathbb{I}_M$: the modes are
quantum-mechanically independent and each saturates Eq.~(\ref{eq:qcrb}).
Crucially, the multi-parameter QCRB is jointly achievable here. The
compatibility (mean Uhlmann curvature) condition for simultaneous saturation
\citep{ParisQMetro2009},
\begin{equation}
\mathcal{U}_{ij} \;\equiv\;
\mathrm{Im}\,\langle\partial_i\psi|\partial_j\psi\rangle
- \mathrm{Im}\big(\langle\partial_i\psi|\psi\rangle\langle\psi|\partial_j\psi\rangle\big)
\;=\;0 ,
\label{eq:uhlmann}
\end{equation}
holds because the generators $\hat Z_i$ are real, commuting multiplication
operators (\ref{app:qfi}). There is therefore \emph{no} quantum
trade-off between simultaneously measured phase modes -- a distinctive feature
of wavefront sensing compared with, e.g., joint phase--amplitude estimation,
where $[\hat Z,\hat n]\neq 0$ enforces incompatibility.
Haffert \etal\cite{Haffert2023} likewise evaluate the pure-state QFI matrix and find it
diagonal, $\mathcal{F}^{Q}=4N_{\mathrm{ph}}\,\mathbb{I}$, for an orthogonal
modal basis; the additional step taken here is the explicit
\emph{saturability} argument. A diagonal QFI matrix bounds each mode
independently but does not, by itself, guarantee that the multi-parameter
bound is \emph{jointly} attainable; the vanishing mean Uhlmann curvature
Eq.~(\ref{eq:uhlmann}) supplies precisely that guarantee, so the
$1/2$~rad-per-photon ceiling is reachable for all low-order modes
\emph{simultaneously}, without mutual quantum penalty.

\subsection{Why intensity-only sensors fall short}
\label{sec:falls}
By the monotonicity of Fisher information under data processing, the
classical FIM of Eq.~(\ref{eq:poissonfim}) obeys $\fim\preceq\mathcal{F}^{Q}$
for every mode: the lantern can at best reach $\beta_i=2$. Intensity-only
focal-plane detection discards the field phase and is a fixed, generally
non-optimal positive-operator-valued measure (POVM); the gap $2-\beta_i$
quantifies the information lost relative to the optimal (interferometric)
measurement that saturates the QCRB. A well-sampled lantern nonetheless
recovers a large fraction of this information for the modes it encodes well,
through inter-port interference of the focal field -- the same mechanism by
which a Zernike-class mask converts phase to intensity
\citep{NDiaye2013,Chambouleyron2023,Haffert2023}.

% =====================================================================
\section{Relation to prior information-theoretic WFS formalisms}
\label{sec:compare}
% =====================================================================
The framework above sits between two recent information-theoretic treatments
of wavefront sensing. Chambouleyron \etal\cite{Chambouleyron2023} model \emph{noise
propagation} for the Fourier-filtering class and define operational
sensitivity metrics; Haffert \etal\cite{Haffert2023} identify the \emph{absolute
ceiling} and a sensor that reaches it. This section states the precise
correspondence in both directions.

\subsection{The Fourier-filtering noise-propagation formalism}
\label{sec:cham}
Chambouleyron \etal\cite{Chambouleyron2023} process the detector signal into reduced
intensities $\Delta I(\phi)=I(\phi)/N_{\mathrm{ph}}-I_0$, referenced to the
flat-wavefront pattern $I_0$, and build an interaction matrix
$\mathcal{D}=[\delta I(\phi_1),\dots,\delta I(\phi_N)]$ by push--pull
calibration, so that $\delta I(\phi_i)=\partial(\Delta I)/\partial\phi_i$.
Reconstruction uses the pseudo-inverse $\mathcal{D}^\dagger$, and propagating
photon and read noise through $\mathcal{D}^\dagger$ yields, for mode
$\phi_i$,
\begin{equation}
\sigma_{\phi_i}^2 =
\frac{N_{\mathrm{sap}}\,\sigma_{\mathrm{ron}}^2}{s^2(\phi_i)\,N_{\mathrm{ph}}^2}
+ \frac{1}{s_\gamma^2(\phi_i)\,N_{\mathrm{ph}}},
\label{eq:cham24}
\end{equation}
their Eq.~(24), with a read-out-noise sensitivity
$s(\phi_i)=\sqrt{N_{\mathrm{sap}}}\,\|\delta I(\phi_i)\|_2$ and a photon-noise
sensitivity
\begin{equation}
s_\gamma(\phi_i)=\left\|\,\delta I(\phi_i)\big/\sqrt{I_0}\,\right\|_2 .
\label{eq:sgamma}
\end{equation}
They observe that these metrics \emph{are} Fisher-information quantities
\citep{Plantet2015}, bounded by the Cram\'er--Rao bound with a maximum value
of two, $0\le s,\,s_\gamma\le 2$ \citep{Paterson2008}, and note explicitly
that ``the Cram\'er--Rao bound actually works for any kind of WFS.'' The
present framework is that general statement, specialised to the lantern and
carried through to the estimator-independent CRLB.

The correspondence is exact and instructive. Identifying the reduced
intensities with the normalised port probabilities,
$I_{0,k}=p_k(\bm 0)$ and $\delta I(\phi_i)_k=\partial p_k/\partial a_i$,
Eq.~(\ref{eq:sgamma}) becomes (\ref{app:equiv})
\begin{equation}
s_\gamma^2(\phi_i)=\sum_{k}\frac{1}{p_k}
\left(\frac{\partial p_k}{\partial a_i}\right)^{\!2}
=\tilde F_{ii},
\label{eq:sgamma_is_Fii}
\end{equation}
i.e.\ the photon-noise sensitivity of Chambouleyron \etal\cite{Chambouleyron2023} is exactly the
\emph{diagonal element} of our per-photon Fisher matrix
Eq.~(\ref{eq:perphoton}). The rigorous per-mode sensitivity, by contrast,
uses the diagonal of the \emph{inverse},
$\beta_i=\sqrt{\eta}\,[(\tilde{\fim}^{-1})_{ii}]^{-1/2}$
(Eq.~\ref{eq:betadef}). Since $(\tilde{\fim}^{-1})_{ii}\ge 1/\tilde F_{ii}$
for any positive-definite $\tilde{\fim}$, with equality \emph{iff} row $i$ of
$\tilde{\fim}$ is diagonal, we have the general ordering
\begin{equation}
\beta_i \;\le\; \sqrt{\eta}\;s_\gamma(\phi_i),
\label{eq:ineq}
\end{equation}
with equality only in the absence of inter-mode cross-talk. In words: the
Chambouleyron photon-noise sensitivity is an \emph{optimistic bound} that
implicitly assumes a diagonal information matrix -- an assumption they state
explicitly -- whereas the CRLB accounts for the covariance between modes. For
Fourier-filtering sensors the calibration modes can be chosen close to the
singular vectors of $\mathcal{D}$, so $\tilde{\fim}$ is nearly diagonal and
the gap in Eq.~(\ref{eq:ineq}) is small. For a photonic lantern this is
generically \emph{not} the case: the adiabatic taper mixes each guided input
mode across many outputs, so the columns of the response are strongly
non-orthogonal and $\tilde{\fim}$ has substantial off-diagonal structure.
Equation~(\ref{eq:ineq}) is then a quantitative correction, and the full
matrix inversion of Eq.~(\ref{eq:crlb}) -- not the per-column norm of
Eq.~(\ref{eq:sgamma}) -- is required for a faithful lantern sensitivity.

Two further differences of scope are worth recording. First, the model of
Chambouleyron \etal\cite{Chambouleyron2023} is explicitly linearised about a flat wavefront
(small-phase, fixed reconstructor $\mathcal{D}^\dagger$); the FIM of
Eq.~(\ref{eq:poissonfim}) is defined for any operating point $\avec_0$ and
any $\mu_k(\avec)$, which is what allows the ensemble-averaged treatment of
Sect.~\ref{sec:obs} for a sensor with vanishing first-order response at zero
phase. Second, their read-out-noise sensitivity $s$ is the diagonal of the
read-noise-weighted FIM of Eq.~(\ref{eq:gaussfim}) in the limit
$\sigma_r^2\gg\mu_k$, so the two-term error budget of Eq.~(\ref{eq:cham24}) is
recovered term by term from Eqs.~(\ref{eq:poissonfim})--(\ref{eq:gaussfim}),
with the $N_{\mathrm{ph}}^{-1/2}$ and $N_{\mathrm{ph}}^{-1}$ scalings of
Sect.~\ref{sec:scaling}.

\subsection{The classical/quantum sensitivity ceiling}
\label{sec:haf}
Haffert \etal\cite{Haffert2023} establish that the optimal wavefront-sensing sensitivity
allowed by classical information theory is $1/2$~radian rms per photon, and --
their central result -- that this classical optimum coincides with the
quantum-metrology limit for starlight, so that $1/2$~rad rms per photon is a
true, measurement-independent floor. In the present notation their limit is
exactly the QCRB of Eq.~(\ref{eq:qcrb}): setting $N_{\mathrm{ph}}=1$ gives
$\sigma_a\ge 1/2$~rad, i.e.\ $\beta=2$. The equality of the classical and
quantum limits that they assert is derived here constructively: the pure-state
QFI of a photonic phase mode is $\mathcal{F}_Q=4\Var_{|A|^2}(\hat Z)$
(Eq.~\ref{eq:qfisingle}), whose per-photon value $4$ over a unit-RMS mode
gives $\beta=2$ and is saturated by an interferometric (phase-referenced)
measurement -- precisely the class the ZWFS and its optimised
PIAA-ZWFS variant approach. Haffert \etal\cite{Haffert2023} show that the PIAA-ZWFS
reaches this fundamental limit for spatial frequencies $>1.7$~cycles/pupil on
arbitrary apertures; the aperture-dependence they treat is the same
$\Var_{|A|^2}(\hat Z)$ rescaling noted after Eq.~(\ref{eq:qcrb}).

The present contribution beyond Haffert \etal\cite{Haffert2023} is twofold. (i) They
derive the diagonal pure-state QFI matrix (their Eq.~19), which bounds each
mode independently; the treatment of Sect.~\ref{sec:multiqfi} adds the
joint-\emph{saturability} proof via the vanishing mean Uhlmann curvature
Eq.~(\ref{eq:uhlmann}), establishing that the $1/2$~rad-per-photon ceiling is
attainable for all low-order modes \emph{simultaneously}, with no residual
quantum incompatibility between them.
(ii) Where the PIAA-ZWFS is engineered to \emph{reach} $\beta=2$, the lantern
is a fixed, mode-mixing device whose $\beta_i<2$ is computed, not designed;
the value of the framework is that it returns $\beta_i$ for \emph{any}
response matrix on the same scale, so a lantern, a pyramid, a ZWFS and a
PIAA-ZWFS can be compared directly.

\subsection{Synthesis}
\label{sec:synth}
Table~\ref{tab:compare} aligns the three conventions. All three are Fisher- or
quantum-Fisher-information constructions on a common $0\le\beta\le2$ scale;
they differ in what they hold fixed. Chambouleyron \etal\cite{Chambouleyron2023} fix a linear
reconstructor and report the diagonal information (an achievable-with-that-
reconstructor sensitivity); this work reports the inverse-matrix CRLB
(the best any unbiased estimator can do); Haffert \etal\cite{Haffert2023} report the
quantum ceiling and a device that attains it. For a cross-talk-free sensor the
first two coincide; for a mode-mixing lantern they differ by
Eq.~(\ref{eq:ineq}), and only the CRLB is estimator-independent.

\begin{table}[t]
\centering
\setlength{\tabcolsep}{8pt}
\caption{Correspondence of the three information-theoretic conventions, all
sharing the ceiling $\beta=s_\gamma=2$ ($1/2$~rad rms per photon): the
photon-noise sensitivity $s_\gamma$ of Chambouleyron \etal\cite{Chambouleyron2023}, the CRLB
metric $\beta$ of this work, and the quantum limit of Haffert \etal\cite{Haffert2023}.}
\label{tab:compare}
\begin{tabular}{@{}l c p{6.4cm}@{}}
\hline\hline
Quantity & Information object & Interpretation \\
\hline
$s_\gamma$ & $\sqrt{\tilde F_{ii}}$ & diagonal FIM; fixed reconstructor \\
$\beta$ (this work) & $[(\tilde{\fim}^{-1})_{ii}]^{-1/2}$ & inverse FIM; any
 unbiased estimator \\
$\tfrac12$\,rad/ph & $\mathcal{F}_Q=4$ & quantum ceiling ($\beta{=}2$) \\
\hline
\end{tabular}
\end{table}

\subsection{What is new relative to prior formalisms}
\label{sec:novelty}
It is worth stating plainly which elements of this framework are
\emph{adopted} from Chambouleyron \etal\cite{Chambouleyron2023} and Haffert \etal\cite{Haffert2023} and
which are \emph{new}. The information-theoretic scaffolding -- the CRLB, the
$\beta\le2$ (equivalently $1/2$~rad-per-photon) ceiling, the identity
$s=\sqrt{\text{Fisher}}$, the $N_{\mathrm{ph}}^{-1/2}$ scaling and the
two-noise-term budget, and the ordering $\fim\preceq\mathcal{F}^Q$ -- is
adopted, not claimed. The following points are, to my knowledge, specific to
this work.
\begin{enumerate}
\item \emph{Device and regime.} Both prior formalisms analyse
Fourier-filtering / pupil-plane mask sensors (the pyramid and Zernike classes
in Chambouleyron \etal\cite{Chambouleyron2023}; the ZWFS and PIAA-ZWFS in
Haffert \etal\cite{Haffert2023}). This is the first Fisher/CRLB \emph{and}
quantum-Fisher treatment of the photonic lantern -- a focal-plane, multi-port,
mode-\emph{mixing} waveguide sensor. That distinction drives points
(ii)--(v).
\item \emph{Diagonal vs.\ inverse information, made precise for a mode-mixing
device.} The photon-noise sensitivity of Chambouleyron \etal\cite{Chambouleyron2023} is exactly
the diagonal of the per-photon FIM, $s_\gamma^2=\tilde F_{ii}$
(Eq.~\ref{eq:sgamma_is_Fii}), whereas the estimator-independent precision uses
the diagonal of its \emph{inverse}, giving $\beta_i\le\sqrt{\eta}\,s_\gamma$
(Eq.~\ref{eq:ineq}). They \emph{assume} the information matrix diagonal (no
inter-mode cross-talk); for a lantern, whose taper mixes each input across many
outputs, that assumption fails, so $s_\gamma$ is an optimistic bound and the
full matrix inversion of Eq.~(\ref{eq:crlb}) is mandatory
(\ref{app:equiv}).
\item \emph{Joint saturability of the multi-parameter quantum bound.}
Haffert \etal\cite{Haffert2023} derive the diagonal pure-state QFI matrix ($F_Q=4$,
off-diagonal zero), but a diagonal QFI matrix does not by itself prove the
multi-parameter bound jointly attainable. The vanishing mean Uhlmann curvature
Eq.~(\ref{eq:uhlmann}), following from the real, commuting phase generators,
supplies that proof: all low-order modes reach $1/2$~rad per photon
\emph{simultaneously} with no residual quantum incompatibility
(Sect.~\ref{sec:multiqfi}, \ref{app:qfi}).
\item \emph{A finite-port capacity ceiling.} The single-frame bound
$M_{\mathrm{eff}}\le N-1$ (Sect.~\ref{sec:obs}): a lantern with $N$
single-mode outputs constrains at most $N-1$ modes per exposure, and low-order
sensitivity saturates once the PSF core is sampled, so added ports principally
buy higher-order capacity. This is architecture-specific to a discrete-port
device and has no analogue in the continuous-detector Fourier-sensor
treatments.
\item \emph{The flat-wavefront operating-point degeneracy.} A symmetric,
intensity-only focal-plane lantern has a \emph{vanishing first-order response}
at a flat wavefront (even modes respond quadratically), so the naive per-frame
FIM is singular; I therefore separate the strict single-frame CRLB from the
ensemble-averaged design information
$\bar{\fim}=\E_{\avec_0}[\fim(\avec_0)]$ (Sect.~\ref{sec:obs}). Fourier-
filtering masks give a first-order signal at a flat wavefront and do not
require this distinction.
\item \emph{Two generalisations and a synthesis.} The QFI is written for a
general, non-uniform pupil illumination,
$\mathcal{F}^Q_{ij}=4\,\Cov_{|A|^2}(\hat Z_i,\hat Z_j)$, with the explicit
$\sqrt{\Var_{|A|^2}(\hat Z_i)}$ rescaling of the $\beta=2$ ceiling for
obstructed or apodised apertures (Haffert \etal\cite{Haffert2023} treat arbitrary pupil
\emph{support} at uniform amplitude); and Table~\ref{tab:compare} places the
three metric conventions -- $s_\gamma$, $\beta$, and ``$1/2$~rad per photon''
-- on a single common $0\le\beta\le2$ dictionary.
\end{enumerate}
The through-line is that Chambouleyron \etal\cite{Chambouleyron2023} and Haffert \etal\cite{Haffert2023} set
the sensitivity \emph{limits} for mask-based sensors; the present work carries
the identical machinery to a mode-mixing, finite-port waveguide sensor, where
the cross-talk correction (ii), the port-capacity ceiling (iv) and the
flat-wavefront degeneracy (v) are new and quantitatively consequential, and
adds the multi-parameter saturability proof (iii) on the quantum side.

% =====================================================================
\section{Illustrative application}
\label{sec:illus}
% =====================================================================
To make the scalings concrete I evaluate
Eqs.~(\ref{eq:poissonfim})--(\ref{eq:betadef}) for a representative,
fully differentiable focal-plane lantern (the analysis code is released with
this paper; see the data availability statement). The aperture field of
Eq.~(\ref{eq:apfield}) is propagated to the focal plane by FFT; $N$
single-mode Gaussian receivers on a centred-hexagonal lattice
($N\in\{7,19,37,91\}$) sample the focal field to give the port amplitudes of
Eq.~(\ref{eq:overlap}); derivatives are formed by central finite differences
and assembled into the FIM. This model is designed to expose scaling and
parity structure, \emph{not} to reproduce a specific fabricated device; a
rigorous beam-propagated transfer matrix enters the identical framework in a
companion numerical study and modifies the parity structure discussed below
through inter-modal mixing.

\begin{figure}[t]
\centering
\includegraphics[width=\columnwidth]{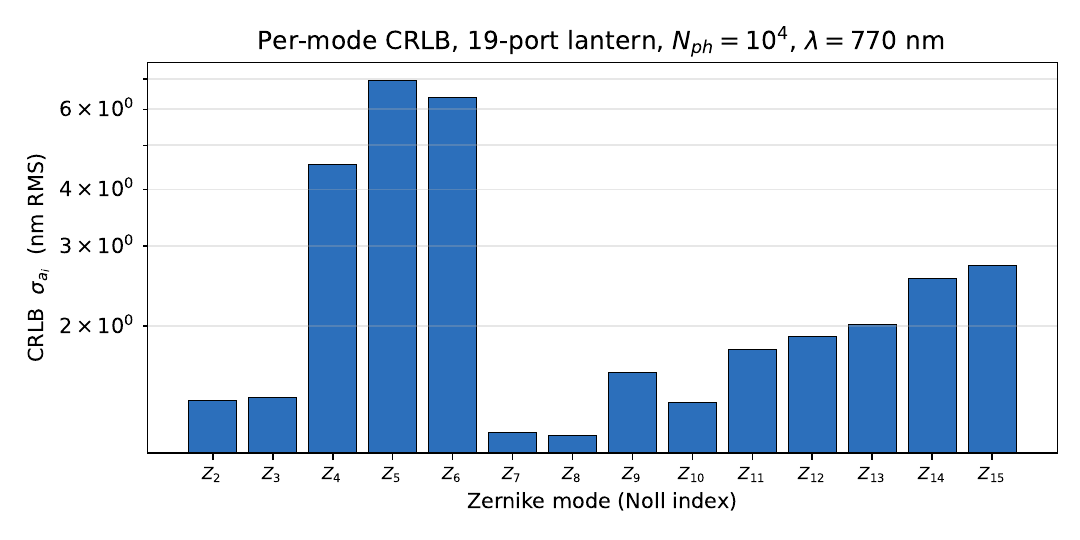}
\caption{Per-mode (averaged) CRLB for a representative 19-port focal-plane
lantern ($N_{\mathrm{ph}}=10^4$). Rotationally-odd modes are sensed near the
information limit; for a symmetric projector the even modes are intrinsically
weaker (Sect.~\ref{sec:illus}).}
\label{fig:permode}
\end{figure}

\begin{figure}[t]
\centering
\includegraphics[width=\columnwidth]{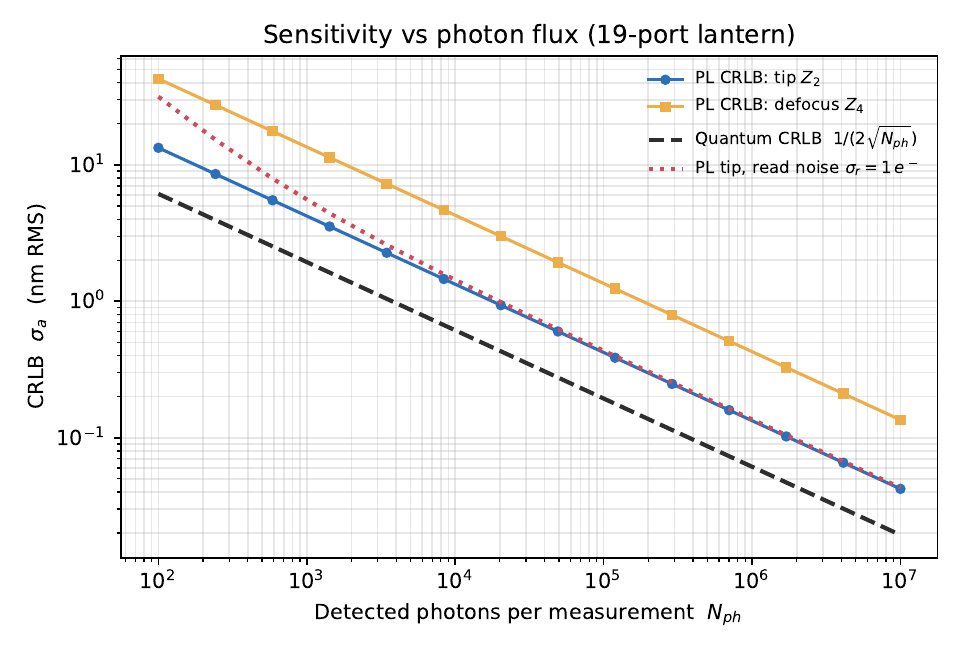}
\caption{CRLB versus detected photons. The photon-limited bound follows
$N_{\mathrm{ph}}^{-1/2}$ (Eq.~\ref{eq:sqrtN}) and sits a fixed factor above the
quantum bound Eq.~(\ref{eq:qcrb}); read noise dominates at low flux, steepening
the slope toward $-1$.}
\label{fig:flux}
\end{figure}

\begin{figure}[t]
\centering
\includegraphics[width=\columnwidth]{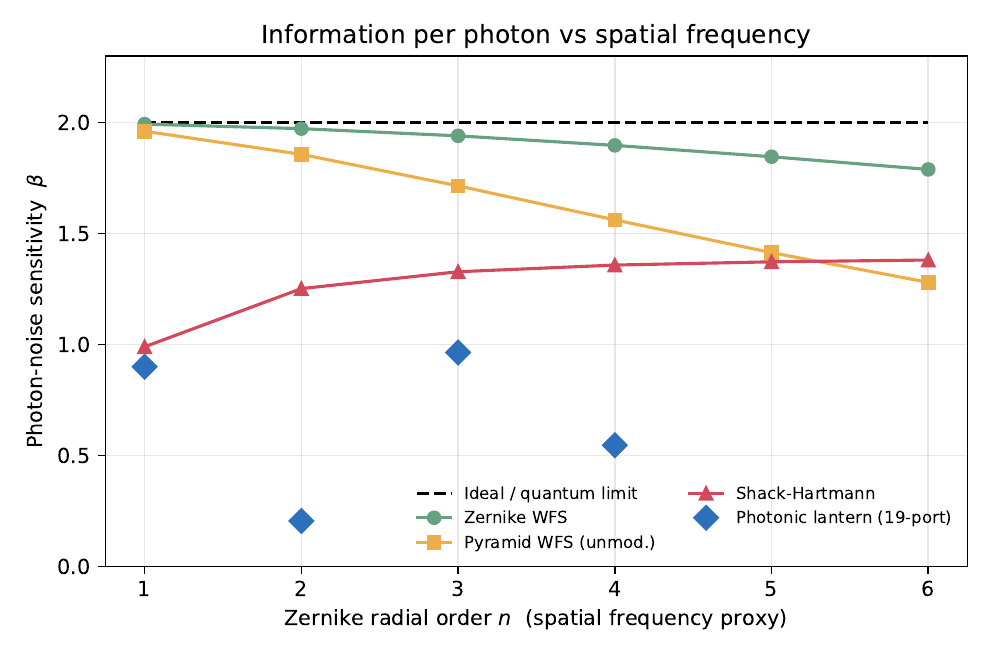}
\caption{Dimensionless sensitivity $\beta_i$ (Eq.~\ref{eq:betadef}) per mode,
referenced to the quantum ceiling $\beta=2$ (Eq.~\ref{eq:qcrb}). This is the
common scale on which the lantern, pyramid, ZWFS and PIAA-ZWFS
\citep{Chambouleyron2023,Haffert2023} can be compared.}
\label{fig:beta}
\end{figure}

\begin{figure}[t]
\centering
\includegraphics[width=\columnwidth]{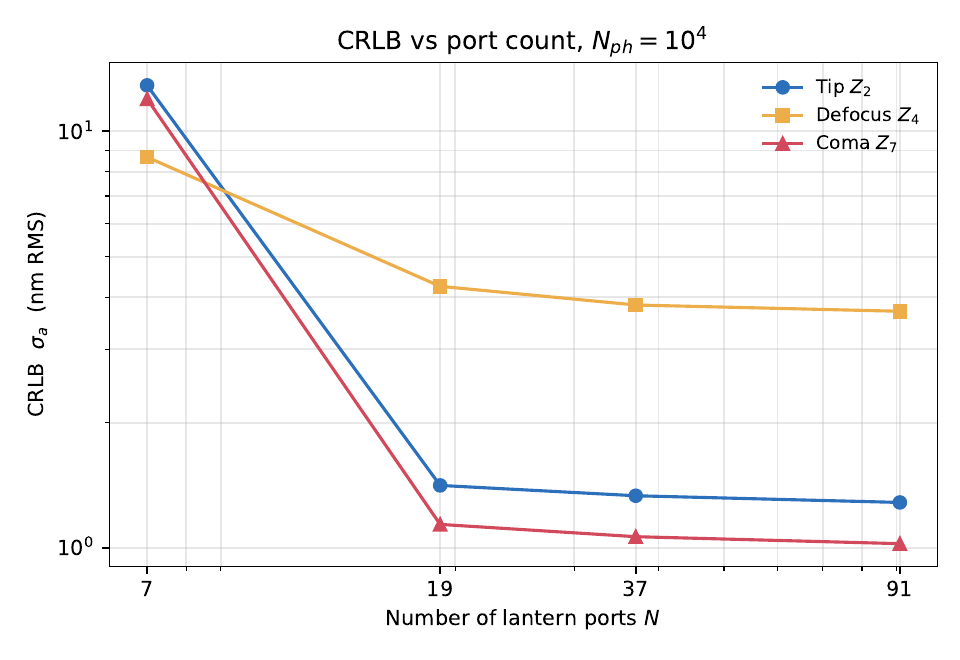}
\caption{Low-order CRLB versus port count at $N_{\mathrm{ph}}=10^4$.
Sensitivity improves sharply from $7$ to $19$ ports and saturates thereafter
for low-order modes -- the per-frame capacity ceiling $M_{\mathrm{eff}}\le N-1$
of Sect.~\ref{sec:obs}.}
\label{fig:ports}
\end{figure}

Figure~\ref{fig:permode} shows the per-mode averaged CRLB for a 19-port
lantern at $N_{\mathrm{ph}}=10^4$: the rotationally-odd modes (tip/tilt, coma,
trefoil) are sensed near the information limit, while the even modes (defocus,
astigmatism) are intrinsically weaker for a \emph{symmetric} focal-plane
projector, whose leading even-mode intensity signature is quadratic. This
parity effect is an estimator-independent property of intensity-only,
symmetric sampling; it is lifted by phase diversity, an asymmetric pupil, or
-- as the companion study shows -- by the intrinsic mode mixing of a real
lantern. Figure~\ref{fig:flux} confirms the $N_{\mathrm{ph}}^{-1/2}$ law of
Eq.~(\ref{eq:sqrtN}) (fitted slope $-0.500$), sitting a fixed factor above the
quantum bound; read noise steepens the curve toward $N_{\mathrm{ph}}^{-1}$ at
low flux, exactly the two regimes of Eq.~(\ref{eq:cham24}).
Figure~\ref{fig:beta} plots $\beta_i$ against the quantum ceiling
$\beta=2$, and Fig.~\ref{fig:ports} shows the low-order CRLB improving sharply
from $7$ to $19$ ports and then saturating -- the quantitative expression of
the per-frame capacity ceiling $M_{\mathrm{eff}}\le N-1$ of
Sect.~\ref{sec:obs}: added ports principally add capacity for higher-order
modes, not precision on modes already well sampled.

% =====================================================================
\section{Conclusions}
\label{sec:concl}
% =====================================================================
I have developed the Fisher-information and Cram\'er--Rao theory of the
photonic-lantern wavefront sensor from the device response, benchmarked it
against the quantum Cram\'er--Rao bound through an explicit multi-parameter
QFI calculation, and placed it in exact correspondence with the two
information-theoretic frameworks that bracket it. The lantern CRLB scales as
$N_{\mathrm{ph}}^{-1/2}$ and is bounded mode-by-mode by the quantum limit
$\sigma_a\ge\lambda/(4\pi\sqrt{N_{\mathrm{ph}}})$, i.e.\ $\beta\le2$. Because
the phase generators are real and commute, the mean Uhlmann curvature vanishes
and the multi-parameter quantum bound is jointly saturable: wavefront sensing
carries no quantum incompatibility between simultaneously estimated modes,
extending the single-mode $1/2$~rad-per-photon ceiling of Haffert \etal\cite{Haffert2023}
to all low-order modes at once. The photon-noise sensitivity $s_\gamma$ of
Chambouleyron \etal\cite{Chambouleyron2023} is exactly the diagonal of the per-photon FIM,
whereas the estimator-independent CRLB uses the diagonal of its inverse; the
two coincide only for a cross-talk-free sensor, so for a mode-mixing lantern
$\beta_i\le\sqrt{\eta}\,s_\gamma$ and the full matrix inversion is required.
The framework is device-agnostic -- it ingests any measured or modelled
response matrix and returns sensitivities on a common $\beta\le2$ scale -- and
thus provides the natural language in which lanterns, pyramid, Zernike and
PIAA-ZWFS sensors can be compared, and against which a beam-propagated lantern
is evaluated in the companion study.

\ack
Partial support for this work was provided by the PICS4SENS project, funded by the
State of Brandenburg through the Investitionsbank des Landes Brandenburg (ILB), with
support from the European Regional Development Fund (ERDF/EFRE), grant number 8600087.

\section*{Data availability statement}
The Fisher/CRLB analysis code that evaluates the framework and reproduces
Figs.~\ref{fig:permode}--\ref{fig:ports} is available from the author (\texttt{code/pl\_crlb\_sim.py}); every routine is annotated with the
equation of this paper that it implements.

\bibliographystyle{iopart-num}
\bibliography{references}

% =====================================================================
\appendix
% =====================================================================
\section{Quantum Fisher information of a photonic phase mode}
\label{app:qfi}

\subsection{Pure-state quantum Fisher information}
Let $\rho_{\avec}=|\psi_{\avec}\rangle\langle\psi_{\avec}|$ be the (pure)
density operator of a single photon in the transverse field, with
$|\psi_{\avec}\rangle=U(\avec)|\psi_0\rangle$ and
$U(\avec)=\exp[-i\sum_i a_i\hat Z_i]$. The QFI matrix is defined through the
symmetric logarithmic derivatives (SLD) $\hat L_i$,
\begin{equation}
\partial_i \rho_{\avec}
= \tfrac{1}{2}\big(\hat L_i\,\rho_{\avec} + \rho_{\avec}\,\hat L_i\big),
\qquad
\mathcal{F}^{Q}_{ij}
= \tfrac{1}{2}\,\Tr\!\big[\rho_{\avec}\{\hat L_i,\hat L_j\}\big],
\label{eq:sld}
\end{equation}
with $\{\cdot,\cdot\}$ the anticommutator. For a pure state the SLD has the
closed form $\hat L_i = 2\,\partial_i\rho_{\avec}
= 2\big(|\partial_i\psi\rangle\langle\psi| + |\psi\rangle\langle\partial_i\psi|\big)$,
and substitution into Eq.~(\ref{eq:sld}) yields the standard pure-state result
\citep{BraunsteinCaves1994,ParisQMetro2009}
\begin{equation}
\mathcal{F}^{Q}_{ij}
= 4\,\mathrm{Re}\!\left[
\langle\partial_i\psi|\partial_j\psi\rangle
- \langle\partial_i\psi|\psi\rangle\langle\psi|\partial_j\psi\rangle
\right].
\label{eq:qfipure}
\end{equation}

\subsection{Phase generators}
Evaluating at $\avec=\bm{0}$, the derivative of the state is
$|\partial_i\psi\rangle = -i\,\hat Z_i|\psi_0\rangle$, since the $\hat Z_i$
commute. Hence
\begin{align}
\langle\partial_i\psi|\partial_j\psi\rangle
   &= \langle\psi_0|\hat Z_i\hat Z_j|\psi_0\rangle
    \equiv \langle \hat Z_i \hat Z_j\rangle, \\
\langle\partial_i\psi|\psi\rangle
   &= +i\langle \hat Z_i\rangle,\qquad
\langle\psi|\partial_j\psi\rangle
    = -i\langle \hat Z_j\rangle .
\end{align}
The moments are intensity-weighted spatial averages,
$\langle \hat Z_i\hat Z_j\rangle=\int |A(\bm r)|^2 Z_i(\bm r)Z_j(\bm r)\dd^2 r$,
because $\hat Z_i$ acts by multiplication and $|\psi_0\rangle$ has transverse
probability density $|A|^2$ (normalised to unity).

\subsection{QFI matrix and its real and imaginary parts}
Inserting the above into Eq.~(\ref{eq:qfipure}), the product
$\langle\partial_i\psi|\psi\rangle\langle\psi|\partial_j\psi\rangle
= \langle \hat Z_i\rangle\langle \hat Z_j\rangle$ is real, and
$\langle \hat Z_i\hat Z_j\rangle$ is real because the $Z_i$ are real
functions. Therefore
\begin{equation}
\mathcal{F}^{Q}_{ij}
= 4\big(\langle \hat Z_i\hat Z_j\rangle
      - \langle \hat Z_i\rangle\langle \hat Z_j\rangle\big)
= 4\,\Cov_{|A|^2}(\hat Z_i,\hat Z_j),
\label{eq:qficov}
\end{equation}
which is Eq.~(\ref{eq:qfimatrix}); the diagonal $i=j$ gives the single-mode
result Eq.~(\ref{eq:qfisingle}). For orthonormal, zero-mean Zernike modes over
a uniformly illuminated pupil $\Cov(\hat Z_i,\hat Z_j)=\delta_{ij}$ and
$\mathcal{F}^{Q}=4\,\mathbb{I}_M$ per photon.

\subsection{Saturability of the multi-parameter bound}
The single-parameter QCRB $\sigma_a^2\ge 1/\mathcal{F}^Q$ is always attainable
by the projective measurement onto the eigenbasis of the SLD. For multiple
parameters the QCRB $\Cov(\hat{\avec})\succeq (\mathcal{F}^{Q})^{-1}$ is
jointly attainable provided the mean Uhlmann curvature vanishes,
$\mathcal{U}_{ij}=\mathrm{Im}\,[\,\langle\partial_i\psi|\partial_j\psi\rangle
-\langle\partial_i\psi|\psi\rangle\langle\psi|\partial_j\psi\rangle\,]=0$
\citep{Helstrom1976,ParisQMetro2009}. From the expressions above both
bracketed terms are real, so $\mathcal{U}_{ij}=0$ identically: the commuting,
real phase generators carry no quantum incompatibility, and all $M$ modes are
estimable simultaneously at the single-mode limit
$\sigma_{a_i}\ge 1/(2\sqrt{N_{\mathrm{ph}}})$.

\subsection{Additivity and the photon-counting limit}
For $N_{\mathrm{ph}}$ independent, identically prepared photons the total
state is a product and the QFI is additive,
$\mathcal{F}^{Q}_{\mathrm{tot}}=N_{\mathrm{ph}}\,\mathcal{F}^{Q}=
4N_{\mathrm{ph}}\,\mathbb{I}_M$, giving the QCRB of Eq.~(\ref{eq:qcrb}).
Finally, the classical FIM of any fixed measurement (here port-intensity
detection, Eq.~\ref{eq:poissonfim}) satisfies
$\fim\preceq\mathcal{F}^{Q}_{\mathrm{tot}}$ by the data-processing inequality,
so $\beta_i\le 2$ for every mode, with equality only for a measurement that
diagonalises the SLD -- an interferometric, phase-referenced detection that
intensity-only sensors do not implement.

% =====================================================================
\section{Equivalence of the photon-noise sensitivity and the Fisher diagonal}
\label{app:equiv}
Working around a flat wavefront, Chambouleyron \etal\cite{Chambouleyron2023} reference the
signal to $I_0$, the flat-wavefront normalised intensity pattern, which in the
present notation is the port probability vector at zero aberration,
$I_{0,k}=p_k(\bm 0)$. Their reduced-intensity derivative is, column by column,
$\delta I(\phi_i)_k=\partial(\Delta I)_k/\partial a_i=\partial p_k/\partial a_i$
(the total-flux normalisation removes any common-mode response). Substituting
into their photon-noise sensitivity Eq.~(\ref{eq:sgamma}),
\begin{equation}
s_\gamma^2(\phi_i)
=\left\|\frac{\delta I(\phi_i)}{\sqrt{I_0}}\right\|_2^2
=\sum_k \frac{1}{p_k}\left(\frac{\partial p_k}{\partial a_i}\right)^{\!2}
=\tilde F_{ii},
\end{equation}
the $i$-th diagonal element of the per-photon Fisher matrix
Eq.~(\ref{eq:perphoton}). The corresponding CRLB variance is
$(\tilde{\fim}^{-1})_{ii}$. For any symmetric positive-definite matrix the
Cauchy--Schwarz (or Kantorovich) inequality gives
$(\tilde{\fim}^{-1})_{ii}\,\tilde F_{ii}\ge 1$, with equality iff the $i$-th
row and column of $\tilde{\fim}$ are diagonal. Hence
$\beta_i=\sqrt{\eta}\,[(\tilde{\fim}^{-1})_{ii}]^{-1/2}\le\sqrt{\eta}\,
\sqrt{\tilde F_{ii}}=\sqrt{\eta}\,s_\gamma(\phi_i)$, i.e.\
Eq.~(\ref{eq:ineq}). The photon-noise sensitivity is thus the CRLB only when
inter-mode information cross-talk is absent; otherwise it overestimates the
achievable precision, and by an amount that grows with the off-diagonal
content of $\tilde{\fim}$ -- which for a mode-mixing lantern is significant.

\end{document}